\documentclass[chicago,apj]{emulateapj}

\slugcomment{}

\shorttitle{CME ENERGY BUDGET}
\shortauthors{MURPHY ET AL.}

\newcommand{\lya}{{\rm Ly}{$\alpha$}}
\newcommand{\lyb}{{\rm Ly}{$\beta$}}
\newcommand{\cii}{{\rm C}~{\sc ii}}
\newcommand{\ciii}{{\rm C}~{\sc iii}}
\newcommand{\hi}{{\rm H}~{\sc i}}

\newcommand{\heii}{{\rm He}~{\sc ii}}
\newcommand{\oi}{{\rm O}~{\sc i}}
\newcommand{\ov}{{\rm O}~{\sc v}}
\newcommand{\ovi}{{\rm O}~{\sc vi}}
\newcommand{\ovia}{{\rm O}~{\sc vi} $\lambda$1031.91}
\newcommand{\ovib}{{\rm O}~{\sc vi} $\lambda$1037.61}
\newcommand{\niii}{{\rm N}~{\sc iii}}
\newcommand{\niiia}{{\rm N}~{\sc iii} $\lambda$990}
\newcommand{\niiib}{{\rm N}~{\sc iii} $\lambda$992}

\newcommand{\ciia}{{\rm C}~{\sc ii} $\lambda$1036.3}
\newcommand{\ciib}{{\rm C}~{\sc ii} $\lambda$1037.0}

\newcommand{\nickeli}{{\rm Ni}~{\sc i}}

\newcommand{\oiv}{{\rm O}~{\sc iv}}
\newcommand{\mgvii}{{\rm Mg}~{\sc vii}}
\newcommand{\feviii}{{\rm Fe}~{\sc viii}}

\newcommand{\blobA}{A} 
\newcommand{\blobB}{B} 
\newcommand{\blobC}{C} 
\newcommand{\blobD}{D} 
\newcommand{\blobE}{E} 
\newcommand{\blobF}{F} 

\newcommand{\kms}{km s\ensuremath{^{-1}}}
\newcommand{\kmss}{km s\ensuremath{^{-2}}}
\newcommand{\cc}{cm\ensuremath{^{-3}}}
\newcommand{\ergg}{ergs~g\ensuremath{^{-1}}}

\newcommand{\vlos}{\ensuremath{V_{\mathrm{LOS}}}}
\newcommand{\vpos}{\ensuremath{V_{\mathrm{POS}}}}

\newcommand{\SOHO}{\emph{SOHO}}
\newcommand{\Yohkoh}{\emph{Yohkoh}}
\newcommand{\Hinode}{\emph{Hinode}}
\newcommand{\GOES}{\emph{GOES}}
\newcommand{\TRACE}{\emph{TRACE}}

\newcommand{\insitu}{\emph{in situ}}
\newcommand{\Insitu}{\emph{In situ}}

\newcommand{\dif}{\ensuremath{\mathrm{d}}}

\begin{document}
  
\title{PLASMA HEATING DURING A CORONAL MASS EJECTION OBSERVED BY
  \SOHO}
\author{N. A. Murphy, J. C. Raymond, and K. E. Korreck}
\affil{Harvard-Smithsonian Center for Astrophysics, 60 Garden Street,
  Cambridge, Massachusetts 02138}
\email{namurphy@cfa.harvard.edu}

\begin{abstract}
We perform a time-dependent ionization analysis to constrain plasma
heating requirements during a fast partial halo coronal mass ejection
(CME) observed on 2000 June 28 by the Ultraviolet Coronagraph
Spectrometer (UVCS) aboard the \emph{Solar and Heliospheric
Observatory} (\emph{SOHO}).  We use two methods to derive densities
from the UVCS measurements, including a density sensitive O V line
ratio at 1213.85 and 1218.35 \AA, and radiative pumping of the O VI
$\lambda\lambda$1032,1038 doublet by chromospheric emission lines.
The most strongly constrained feature shows cumulative plasma heating
comparable to or greater than the kinetic energy, while features
observed earlier during the event show cumulative plasma heating
of order or less than the kinetic energy.  \emph{SOHO} Michelson
Doppler Imager (MDI) observations are used to estimate the active
region magnetic energy.  We consider candidate plasma heating
mechanisms and provide constraints when possible.  Because this CME
was associated with a relatively weak flare, the contribution by flare
energy (e.g., through thermal conduction or energetic particles) is
probably small; however, the flare may have been partially behind the
limb.  Wave heating by photospheric motions requires heating rates
significantly larger than those previously inferred for coronal holes,
but the eruption itself could drive waves which heat the plasma.
Heating by small-scale reconnection in the flux rope or by the CME
current sheet is not significantly constrained.  UVCS line widths
suggest that turbulence must be replenished continually and dissipated
on time scales shorter than the propagation time in order to be an
intermediate step in CME heating.
\end{abstract}


\section{INTRODUCTION\label{introduction}}
 
The understanding of astrophysical phenomena generally begins with the
energy budget.  The energy budgets of coronal mass ejections (CMEs)
include contributions from magnetic energy, bulk kinetic energy,
ionization energy, gravitational potential energy, thermal energy, and
energetic particles \citep[e.g.,][]{emslie:2005}.  Energy can be lost
to the system through radiation and thermal conduction.  White light
observations from instruments such as the Large Angle Spectroscopic
Coronagraph \citep[LASCO;][]{brueckner:lasco:1995} on board the
\emph{Solar and Heliospheric Observatory} (\SOHO) allow a
straightforward determination of the contributions by kinetic and
gravitational energy \citep[e.g.,][]{vourlidas:2000, vourlidas:2010,
subramanian:2007}.  The magnetic energy is thought to be the largest
component of the CME energy budget but is difficult to constrain via
remote sensing because of the lack of good magnetic field diagnostics,
especially for transient events such as CMEs\@.  However, the total
and free magnetic energy of precursor active regions can be estimated
using vector magnetograms \citep[e.g.,][]{metcalf:2005}, line-of-sight
magnetograms from the Michelson Doppler Imager (MDI) on \SOHO, and
empirical relationships between X-ray luminosity and magnetic flux
\citep{fis98}.  The magnetic field structure of interplanetary coronal
mass ejections (ICMEs) can be investigated using \insitu\ measurements
\citep[see][and references therein]{zurbuchen:2006} by spacecraft such
as the \emph{Advanced Composition Explorer} (\emph{ACE}).

Several different lines of evidence now suggest that the thermal
energy input into CMEs is comparable to the kinetic energy of the
ejected plasma.  First, observations between 1.5 and 3.5 solar radii
by the Ultraviolet Coronagraph Spectrometer
\citep[UVCS;][]{kohl:uvcs:1995, kohl:2006} on board \SOHO\ have been
analyzed using a time-dependent ionization code for four prior events
\citep{akmal:2001, ciaravella:2001, lee:2009, landi:2010}.  The
thermal energy input was constrained to be comparable to or greater
than the kinetic energy for several features during each of these
events.  This method is used in this paper and further described in
Section \ref{method}.  Second, \insitu\ measurements of ICMEs
typically show high ion charge states \citep[e.g.,][]{lepri:2001,
lynch:2003, lepri:2004}, although low charge state plasma is sometimes
observed \citep[e.g.,][]{lepri:2010}.  Detailed models of ICMEs by
\citet{rakowski:2007} showed that continual heating was required out
to several solar radii to be consistent with the observed charge
states of both iron and oxygen at 1 AU, with total heating
requirements again found to be comparable to the CME kinetic energy
\citep[see also][]{gruesbeck:2011}.  Third, \citet{filippov:2002}
presented observations of a rising prominence by the \emph{Transition
Region and Coronal Explorer} (\TRACE).  During the eruption, the
feature seen at 171 \AA\ suddenly changed from absorption to emission
by the prominence, indicating significant and rapid heating.  Fourth,
\citet{liu:2006} investigated \insitu\ measurements between $0.3$ and
$20$~AU and found that dissipation of turbulence can explain the
observed heating rates.  The turbulence generation mechanism is not
understood in the inner heliosphere but heating by pickup ions
contribute to ICME heating in the outer heliosphere.

Several analytical models explore the energetics of expanding flux
ropes.  \citet[][hereafter, KR]{kumar:rust:1996} assume global
conservation of mass, magnetic flux, and helicity for a self-similarly
expanding force-free flux rope.  The magnetic energy is found to
decrease monotonically during the process of expansion.  Some of the
lost magnetic energy goes into overcoming solar gravity and increasing
the bulk kinetic energy, but a large fraction is presumed to go into
heating the plasma within the expanding flux rope.  \citet{wang:2009}
develop a different model of a self-similarly expanding flux rope
which relaxes several assumptions made by KR and implicitly includes
some effects associated with the solar wind.  Their inference of a CME
polytropic index of $\frac{4}{3}$ for one event indicates continued
heating during flux rope expansion.  \citet{lyutikov:2011} model CME
flux ropes as expanding force-free spheromaks.  By allowing for finite
dissipation during expansion and assuming a large anomalous
resistivity (e.g., due to wave-particle interactions),
\citet{rakowski:2011} show that this model can reproduce charge states
observed within the flux rope, but not the higher charge states such
as {\rm Fe}$^{16+}$ in the plasma trailing the flux rope.

However, none of these models specify the physical mechanisms
responsible for plasma heating.  The most likely candidate is
dissipation of magnetic energy.  There are several candidate
mechanisms, which are discussed in detail in Section \ref{mechanisms}.
These include upflow from the CME current sheet, kinking of the CME
flux rope, small-scale reconnection or tearing behavior within the CME
ejecta, waves driven by either photospheric motions or by the eruption
itself, thermal conduction, and deposition of energy by energetic
particles.  Additionally, \citet{filippov:2002} argue that colliding
flows along flux tubes can heat prominence plasma to coronal
temperatures in upward concave regions of flux tubes.

In this paper we use a time-dependent ionization analysis to constrain
the thermal energy content and plasma heating of a fast CME observed
by \SOHO\ on 2000 June 28\@.  This technique has previously been used
to investigate heating during four slow CMEs observed by \SOHO/UVCS
(see Table \ref{census}).
\citet{akmal:2001} analyzed a CME on 1999 April 23 and found that the
cumulative heating energy was comparable to the kinetic and
gravitational potential energy of this event.  There was a core of
cool plasma radiating in \ciii\ which was surrounded by hotter
material radiating in \ovi\@.
\citet{ciaravella:2001} analyzed a CME on 1997 December 2 and found
that gradual heat release mechanisms are probably more appropriate than
mechanisms where heating is concentrated during the early evolution of
a CME\@.
\citet{lee:2009} investigated a CME observed on 2001 December 13 and
constrained the cumulative heating energy for the bright knots
observed in \ovi\ to be greater than the kinetic energy.
\citet{landi:2010} analyzed the `Cartwheel CME' observed by \SOHO,
\emph{Hinode}, and \emph{STEREO} on 2008 April 9 and again found that
the cumulative heating energy was constrained to be greater than the
kinetic energy.  
Both \citet{lee:2009} and \citet{landi:2010} show that thermal
conduction is probably too slow to sufficiently heat the CME plasma
during the early evolution of the flux rope.
Using different methods, \citet{bemporad:2007} investigated a slow
flareless CME observed by UVCS and the Large Angle and Spectrometric
Coronagraph Experiment (LASCO) on 2000 January 31 and found that the
core and leading edge of the CME were hotter than the ambient corona.

\begin{deluxetable}{cccc}
\tabletypesize{\scriptsize}
\tablecaption{Census of CME energy budget papers using the technique
  of \citet{akmal:2001}\label{census}}
\tablewidth{0pt}
\tablehead{
\colhead{Event} &
\colhead{\vpos\ (km/s)} &
\colhead{Mass (g)} &
\colhead{Reference} 
}
\startdata
1997 Dec 12 & $211$  & $2.9\times 10^{14}$  &
\citet{ciaravella:2001} 
\\
1999 Apr 23 & $523$  & $2.7\times 10^{15}$  & \citet{akmal:2001}
\\
2000 Jun 28\tablenotemark{a} & $1198$  & $7.3\times 10^{15}$ & Murphy et al.\
(2011) 
\\
2001 Dec 13\tablenotemark{a} & $864$  & --- & \citet{lee:2009} 
\\
2008 Apr 9  & $650$ & $1.4\times 10^{15}$  & \citet{landi:2010}
\enddata
\tablenotetext{a}{Partial halo CME; uncertain mass.}
\end{deluxetable}

The 2000 June 28 CME analyzed in this paper has been previously
investigated in multiple separate works.
\citet{raymond:2004} used radiative pumping of the \ovi\
$\lambda\lambda 1032, 1038$ doublet observed by UVCS to find number
densities ranging from {$\sim$}{$1.28\times 10^6$}--$4\times 10^7$
{\cc}.  This method is described further in Section \ref{method} and is
used in this work to find densities for several features along the
UVCS slit.
\citet{ciaravella:2005} used UVCS and LASCO observations to identify
the leading edge of this CME as a fast mode shock front.
\citet{maricic:2006} studied the kinematics of the prominence and
leading edge of this CME with LASCO and the Mark IV (MK4) coronagraph
at the Mauna Loa Solar Observatory (MLSO)\@.
This CME has also been identified as a solar energetic particle (SEP)
event \citep{ho:2003, wang:2006}.

The organization of this paper is as follows.  
In Section \ref{diagnostics}, we discuss the methods we use to find
densities using UVCS spectra.
Observations of the 2000 June 28 CME are described in Section
\ref{dataset}.
Section \ref{features} identifies UVCS features that have good density
diagnostics.
The method for our time-dependendent ionization analysis of this event
is described in Section \ref{method}.
A discussion of the results of this analysis is presented in Section
\ref{results}.  
The magnetic energy of the precursor active region is estimated in
Section \ref{magneticenergy}.  
Constraints on candidate heating mechanisms are discussed in Section
\ref{mechanisms}.
Section \ref{conclusions} contains a summary and conclusions.  

\section{DENSITY DIAGNOSTICS}
\label{diagnostics}
A key component of the time-dependent ionization analysis performed in
this paper is the electron number density at the location in the
corona observed by UVCS\@.
Knowledge of the UVCS density allows us to exclude a significant
portion of parameter space when using the method described in Section
\ref{method}.
In this section we discuss our two primary density
diagnostics: a classical density sensitive \ov\ line ratio
\citep{akmal:2001} and radiative pumping of the \ovi\ doublet by
chromospheric lines \citep{raymond:2004}.  

\subsection{The density-sensitive $\lambda\lambda 1213,1218$ \ov\ line 
  ratio}

\begin{figure}
  \begin{center}
  \includegraphics{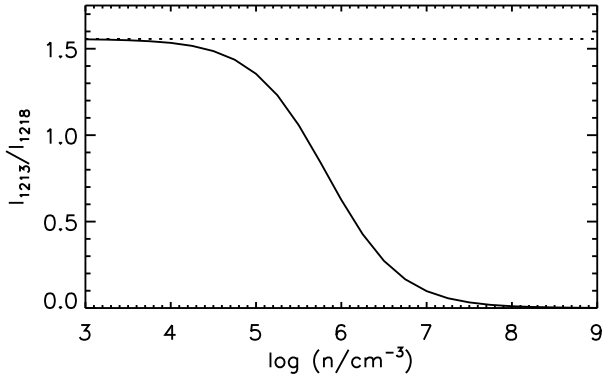}
  \end{center}
  \caption{The ratio of the [\ov] line at 1213.85 \AA\ to the \ov] line
  at 1218.67 \AA\ as a function of electron density at $2.24\times
  10^5$~K\@.\label{ovplot}}
\end{figure}

The ratio of the [\ov] forbidden line at 1213.85 \AA\ to the \ov]
intercombination line at 1218.35 \AA\ is a useful diagnostic for
coronal densities of {$\sim$}{$10^6$} \cc\ \citep{akmal:2001} and is
based on well-understood atomic physics.  At high densities the
forbidden line is weak because of collisional deexcitation.  These
lines occur at equilibrium temperatures of $\sim${$2\times 10^5$}~K.
We use the CHIANTI database \citep{dere:1997,dere:2009}, which
includes proton collisions.  The intensity ratio of these two lines,
$I_{1213}/I_{1218}$, is shown as a function of electron number density
in Figure \ref{ovplot}.

The [\ov] and \ov] lines straddle \lya\ at 1215.67 \AA\@.  Bright
\lya\ emission can obscure weak [\ov] emission to make it difficult to
get an accurate line strength for \ov]\@.  Moreover, because of the
construction of UVCS (described in Section \ref{uvcs_section}), the
secondary channel [\ov] line sometimes appears at the same location on
the detector as the primary channel \niii\ $\lambda 991.58$ \AA\ line.
Therefore, this density diagnostic is most useful at certain
instrument-dependent line-of-sight velocities and when \lya\ emission
is relatively weak, but has the advantage that it is weakly dependent
on temperature.

\subsection{Radiative pumping of O VI}

The second density diagnostic is the intensity ratio of \ovi\ $\lambda
1031.91$ to \ovi\ $\lambda 1037.61$.  When collisional excitation
dominates, the ratio will be 2:1.  Departures from this ratio are due
to radiative pumping of \ovi\ $\lambda 1037$ by \cii\ $\lambda\lambda
1036.3$, $1037.0$ near velocities of 172 and 371 km s$^{-1}$, \ovi\
$\lambda 1037$ by \ovi\ $\lambda 1032$ near velocities of 1650 km
s$^{-1}$, or \ovi\ $\lambda 1032$ by {\lyb} near velocities of 1810 km
s$^{-1}$ \citep{noci:1987, raymond:2004}.

The electron number densities are given as follows.  
For pumping of \ovi\ $\lambda 1037.61$ by \cii\ $\lambda1036.3$ near
velocities of 371 km s$^{-1}$ \citep[see also][]{noci:1987}, the ratio
is less than 2:1 and the number density is given by
\begin{equation}
  n_e = \frac{\sigma_{1037}I_{\mathrm{disk}}(\mbox{\cii})W}{q_{1037}}
        \frac{2-R}{2R}.
\end{equation}
For pumping of \ovi\ $\lambda 1037.61$ by \ovi\ $\lambda 1031.91$ near
velocities of 1650 km s$^{-1}$, the ratio is less than 2:1 and the
number density is given by
\begin{equation}
  n_e = \frac{\sigma_{1037}I_{\mathrm{disk}}(1032)W}{q_{1037}}
        \frac{2-R}{2R}.
\end{equation}
For pumping of \ovi\ $\lambda 1031.91$ by \lyb\ near velocities of
1810 km s$^{-1}$, the ratio is greater than 2:1 and the number density
is given by
\begin{equation}
  n_e = \frac{\sigma_{1032}I_{\mathrm{disk}}(\mbox{\lyb})W}{q_{1032}}
        \frac{2}{R-2}.
\end{equation}
In the above relations, $\sigma_{1032}$ and $\sigma_{1037}$ are the
effective scatting cross sections, $I_{\mathrm{disk}}$ is the solar
disk intensity of each of the lines, $W\equiv 2\pi \left(1 -
\sqrt{1-r^{-2}}\right)$ is the dilution factor for a distance $r$ from
Sun center, $q_{\mathrm{1032}}$ and $q_{\mathrm{1037}}$ are the
collisional excitation rate coefficients, and $R\equiv
I_{1032}/I_{1038}$ is the intensity ratio of the two \ovi\ lines.

This density diagnostic requires more assumptions than the \ov\
density sensitive line ratio.  The caveats of this method are
discussed by \citet{raymond:2004} and include that the solar disk
intensity of each of the illuminating lines may be enhanced early
during flares \citep{raymond:2007, johnson:2011}, multiple components
to the plasma can exist at different speeds along the line of sight,
and the collisional excitation rate depends relatively strongly on the
temperature of the plasma.  In addition, weak \cii\ $\lambda\lambda
1036.3$, $1037.0$ lines complicate the process of finding the \ovi\
line ratio from observational data.  We mitigate the effects of these
\cii\ lines by fitting Gaussian line profiles to \lyb\ and the \ovi\
doublet and enforcing that both lines in the \ovi\ doublet have the
same width.  In Section \ref{features} we compare the densities
derived through both the \ov\ and \ovi\ diagnostics for one particular
feature (blob E) and find that the results are consistent to within
expected systematic uncertainties.

\section{DATA SET}
 
\label{dataset}

Observations of the CME on 2000 June 28 are available from the EIT,
LASCO, and UVCS instruments on board \SOHO, and the MK4 coronagraph at
MLSO\@.  The C class flare associated with this CME was detected by
\GOES\@.  Pre-CME and post-CME observations of this event were made by
\Yohkoh/SXT and \SOHO/MDI\@.  The observations for this event have been
previously described by \citet{ciaravella:2005} and
\citet{maricic:2006}, which we summarize below.

\subsection{\SOHO/EIT observations}
\label{eit}

\begin{figure}
  \begin{center}
  \includegraphics[scale=1.0]{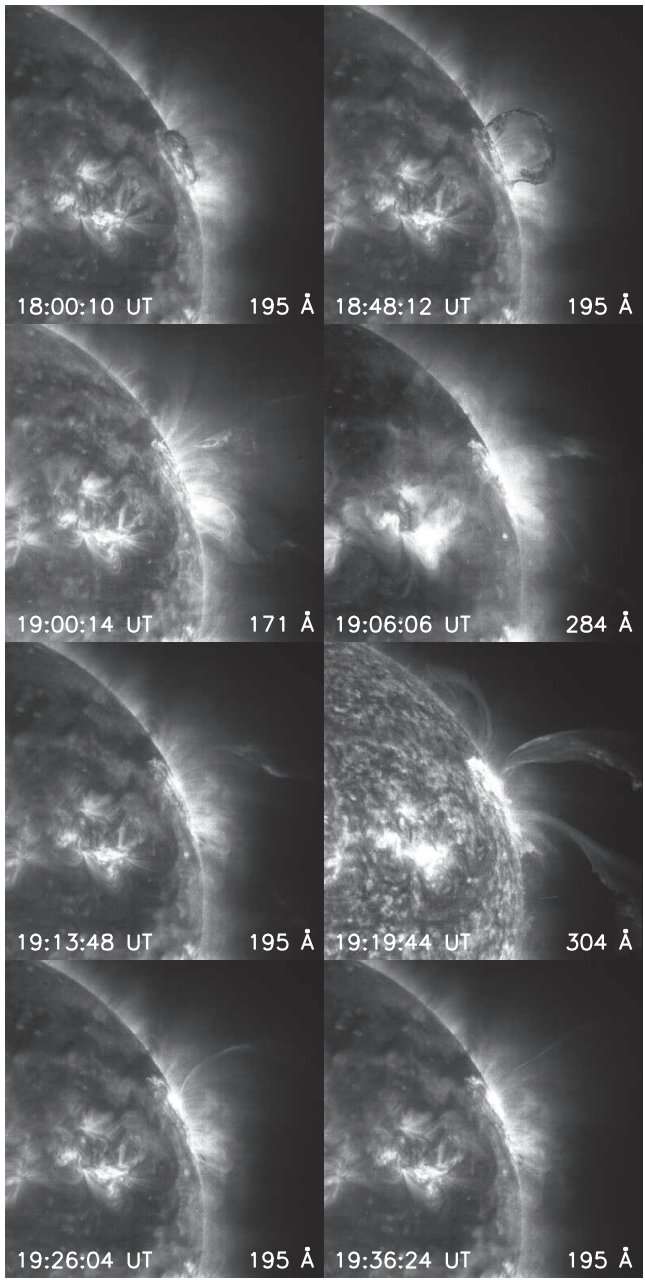}
  \end{center}
  \caption{ \SOHO/EIT observations of the 2000
    June 28 CME, rescaled to emphasize faint structure.  
  \label{eitfig}}
\end{figure}

\SOHO/EIT images in the 195 \AA\ band were taken with a $\sim${$12$}
minute cadence during most of this event.  Between 19:00:14 and
19:19:44 UT, EIT images at 171 \AA, 284 \AA, 195 \AA, and 304 \AA\
were taken in order with a six minute cadence.  Several of these
observations are shown in Figure \ref{eitfig} from 18:00 UT until
19:36 UT \citep[see also][Figure 1]{ciaravella:2005}.  The series of
195 \AA\ observations shows a dark arcade near the northwest limb
starting to rise around 18:00:10 UT\@.  That it is dark at 195 \AA\
indicates the absence of plasma at $\sim$1~MK in the rising structure.
Between 18:36:10 and 18:48:12 UT, this feature rises rapidly and is
not apparent in the next 195 \AA\ observation at 19:13:48 UT\@.  The
304 \AA\ image at 19:19:44 UT shows bright strands indicative of
\heii\ emission, suggesting that the rising filament was relatively
cool.  The 195 \AA\ images at 19:26:04 and 19:36:24 UT show a thin
strand of hot plasma that outlines the \heii\ arch in the 304 \AA\
image at 19:19:44 UT, indicating the presence of some plasma around 1
MK associated with the strand.  These strands are probably associated
with the feature identified in Section \ref{features} as blob F\@.

\subsection{MLSO/MK4 observations}

\begin{figure}
  \begin{center}
  \includegraphics[scale=1]{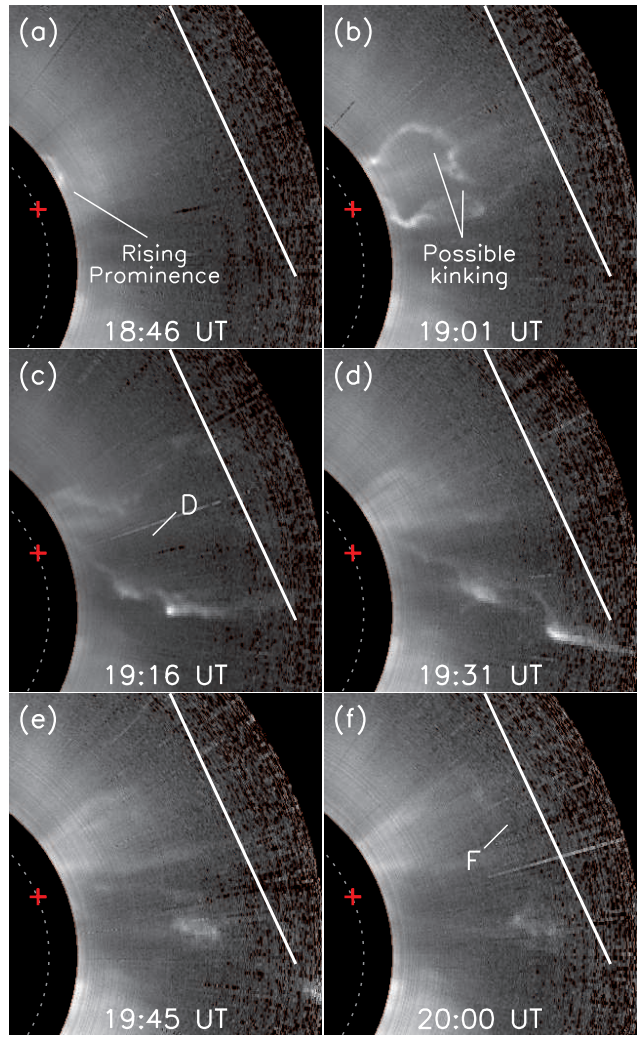}
  \end{center}
  \caption{ MLSO/MK4 polarization brightness observations of the 2000
    June 28 CME, rescaled to emphasize faint structure.  Estimated
    positions of the features observed by UVCS and discussed in
    Section \ref{features} are labeled.  The position of the precursor
    active region is denoted by the red plus sign.
  \label{mlso}}
\end{figure}

MLSO/MK4 measures the polarization brightness (pB) of the corona
between $\sim${$1.1$} and $\sim${$2.8$} solar radii using radial scans
with a cadence of about three minutes.  MK4 observations were
performed on the day of the event between 16:56 and 20:00 UT\@.  A
time sequence of the eruption is presented in Figure \ref{mlso}.

The first clear sign of the CME was at 18:46 UT when a rising bright
arch entered the field of view (see Figure \ref{mlso}a).  We identify
this arch as the erupting filament observed by EIT, most notably at
18:48 UT (see Figure \ref{eitfig}).  Consequently, this
filament is observed simultaneously by EIT and MK4\@.  The kinematics
of this rising prominence were presented by \citet{maricic:2006}.  The
prominence was accelerated most strongly between $\sim$18:40 and
$\sim$19:00 UT at up to $\sim${$0.5$} \kmss\@.  The upper parts of
this arch begin to fade around 19:01 UT (Figure \ref{mlso}b) and are
not apparent by 19:16 UT (Figure \ref{mlso}c).  

After about 18:49 UT, the rising loop developed a twisted or helical
structure.  This behavior is consistent with the onset and nonlinear
growth of a long wavelength kink instability, and is apparent in the
observation at 19:01 UT (Figure \ref{mlso}b).  A bright stream of
ejecta directed towards the southwest is observed to miss the UVCS
slit entirely (Figure \ref{mlso}c--d).  A faint feature rising at
$\vpos\approx$250 \kms\ is observed around 20:00 UT when the set of
observations ends, which we identify in Section \ref{features} as Blob
F\@.

\subsection{\SOHO/LASCO white light observations}

\begin{figure}
  \begin{center}
  \includegraphics[scale=1]{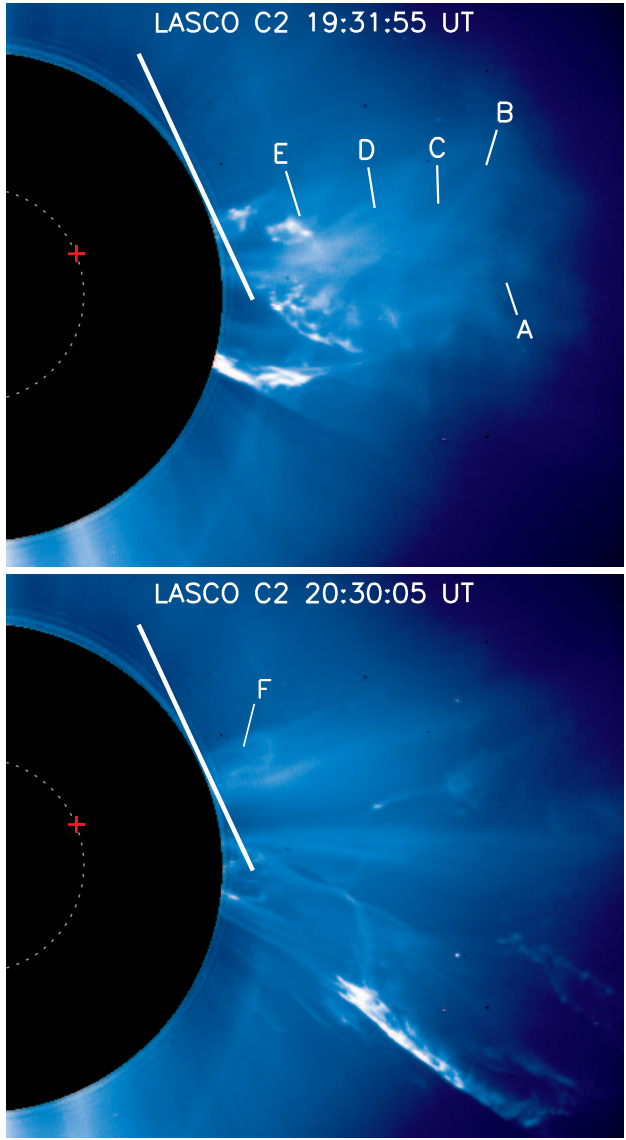}
  \end{center}
  \caption{LASCO C2 white light observations of this event, rescaled
    to emphasize faint structure.  The approximate locations of the
    features observed by UVCS and analyzed using a time-dependent
    ionization code are labeled.  The plane of sky velocities for
    these features were estimated using either \ovi\ pumping
    information or complementary MLSO/MK4 observations.  The precursor
    active region is denoted by the red plus sign.  \label{lascofig}}
\end{figure}

\SOHO/LASCO performs white light observations of the solar corona
between 2.5 and 30 solar radii.  Observations of the 2000 June 28 CME
were performed by both the C2 and C3 cameras for the duration of the
event.  The first C2 observation at 19:31:55 UT of the eruption showed
a large plume of plasma off of the west limb (see Figure
\ref{lascofig}).  The flux rope observed by EIT and MK4 had by this
time broken up into several different outwardly propagating blobs.  
In subsequent C2 observations at 19:54:41 and 20:06:05 UT, each of the
clouds propagated approximately radially outward from the location of
the flare site and rising prominence with some expansion.  The C2
image at 20:30:05 UT shows the loop identified as Blob F in Section
\ref{features} slowly rising above the center of the UVCS slit.  This
feature was probably ejected late during the eruption.

Observations by both the C2 and C3 cameras show a bright stream of
plasma propagating approximately radially outward from the region of
the flare site.  Unfortunately, this feature passed just south of the
UVCS slit.  The feature exhibited significant expansion during
propagation.  The apparent velocity of this feature was {$\sim$}$300$
\kms\ at the low altitude end and $\sim${$600$} \kms\ at the high
altitude end while in the C2 field of view.

The LASCO CME Catalog estimates a mass of {$\sim$}$7.3\times
10^{15}$~g and a kinetic energy of {$\sim$}$5.3\times 10^{31}$ ergs.
This corresponds to a kinetic energy per unit mass of
{$\sim$}$73\times 10^{14}$ \ergg\@.
However, the mass of this event is uncertain because this was a
partial halo CME; furthermore, the kinetic energy is uncertain because
it assumes that the entire plasma is propagating at the same velocity,
and only considers the plane-of-sky velocity.

\subsection{\SOHO/UVCS observations}

\label{uvcs_section}

\SOHO/UVCS \citep{kohl:uvcs:1995, kohl:2006} is a long slit
spectrograph designed to study the solar corona at 1.5--10 solar
radii.  The use of both internal and external occulters keeps stray
light at low enough levels to observe the faint corona.  The UVCS
\ovi\ channel contains two spectrometer light paths optimized for the
\ovi\ $\lambda\lambda 1032,1038$ doublet (primary) and \lya\
(redundant).  Throughout the course of this event, the UVCS entrance
slit was positioned at $\rho=2.32R_{\odot}$ with a position angle of
$295^{\circ}$ over the precursor active region near the northwest
limb.  The exposure time was two minutes, thus allowing high time
resolution observations of the ejecta.  The slit width was $49$
$\mu$m.  The observations are available in three wavelength ranges
(denoted panels): 976--979 \AA\@, which contains the \ciii\
$\lambda$977 line; 1024--1045 \AA, which contains the \ovi\
$\lambda$1032,1038 doublet and \hi\ \lyb; and 1210--1220 \AA, which
contains \hi\ \lya\ and the forbidden and intercombination lines of
\ov\ at 1213 \AA\ and 1218 \AA, respectively.  Because the positions
on the detector overlap between the different channels, the primary
channel \niii\ $\lambda$989,991 doublet appears in the \lya\ panel at
different wavelengths depending on the redshift.  Because of telemetry
limitations, the UVCS observations were binned in three pixels in the
spatial direction for a resolution of $21''$ per bin, and two pixels
in the spectral direction for a resolution of 0.198 \AA\ (57 \kms) for
the \ciii\ and \ovi\ panels and 0.183 \AA\ (45 \kms) in the \lya\
channel.

\begin{figure}
  \begin{center}
  \includegraphics[scale=1.0]{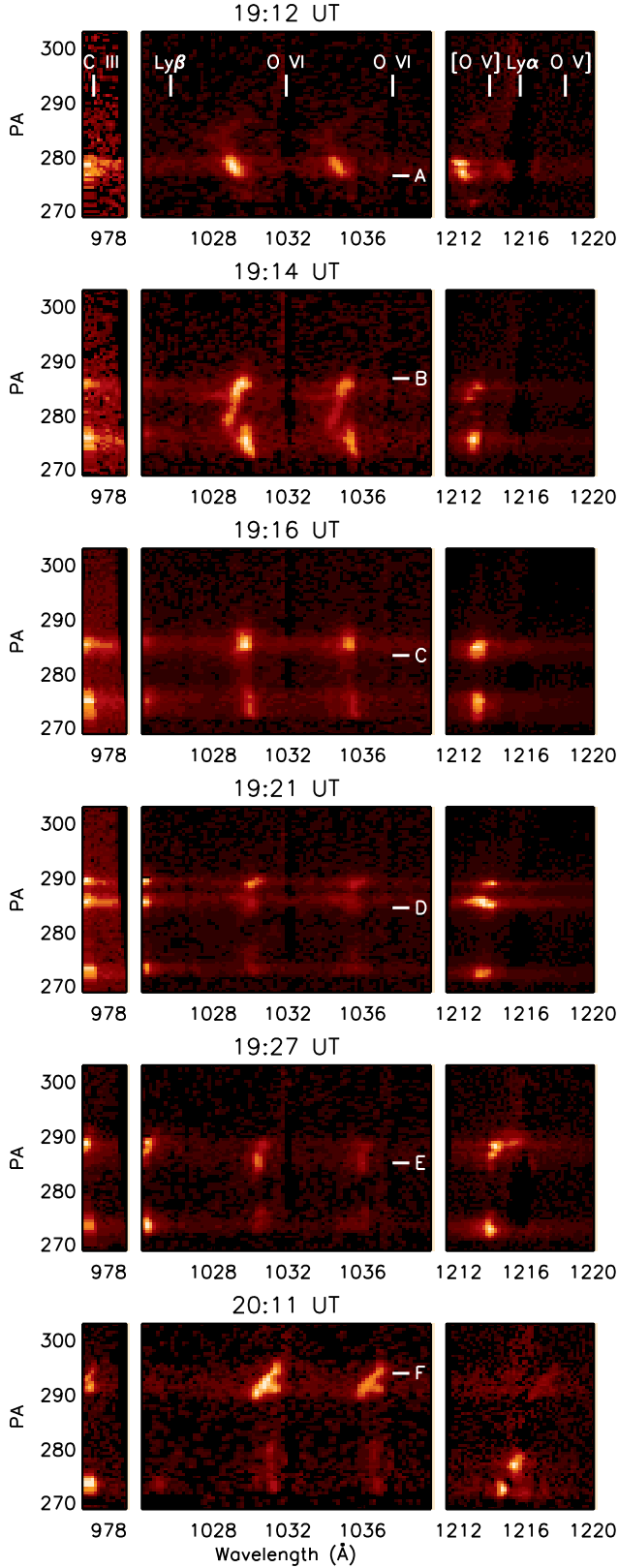}
  \end{center}
  \caption{UVCS observations of each of the features identified in
    Section \ref{features}, rescaled to emphasize faint structure.
    The pre-CME average background has been subtracted.  From left to
    right are the \ciii\ panel, the \ovi\ panel, and the \lya\ panel.
    The rest wavelengths of the features are shown in the top row.
    \label{uvcsfig}}
\end{figure}

The UVCS observations of this event are summarized in detail by
\citet{ciaravella:2005}.  Starting in the exposure at 18:59, UVCS
detected faint, diffuse, and broad blueshifted \ovi\ emission at
position angles of 273--295$^\circ$, indicating passage of the CME
front.  Knots bright in \ovi\ and \lya\ first appear at 19:06 UT at a
position angle of 273$^\circ$.  At 19:12 UT, a blueshifted bright knot
in \ovi\ and \lya\ appears at 277$^\circ$ (see Figure \ref{uvcsfig};
this is associated with blob A).  This indicates the passage of cooler
prominence plasma.  By 19:14 UT, these knots fill the spatial
direction along the slit between $\mathrm{P.A.}=272^\circ$ and
$290^\circ$.  Shortly after, two knots persist in the UVCS
observations but the southern knot fades by about 19:30 UT\@.  For
many of the features early in the event, \lyb, [\ov], and \ciii\ were
off-panel because of the very large blueshifts.  The northern knot
persists for a while but drifts southward.  Between 20:09 and 20:13
UT, a new diagonal or shear flow feature appears between
$\mathrm{P.A.}=288^\circ$--$297^\circ$ that has apparent \ov], \ovi,
and \ciii\ emission but little \lya\ or \lyb\@.  A few plasma streams
persist afterward as the corona settles back into an undisturbed
state.

\subsection{\GOES\ {X}-ray observations}

\label{goesobservations}

A C class flare was observed by \GOES\ starting at 18:48 UT with the
decay continuing until about two hours after the event.  Using the
\GOES\ $1$--$8$ \AA\ light curve and subtracting the background flux,
we estimate that the amount of flare energy emitted in this band is
$E_X \sim 3\times 10^{28}$ ergs over the entire event.  Assuming
$L_{\mathrm{tot}}/L_{\mathrm{X}}=100$ \citep[e.g.,][]{emslie:2005},
this corresponds to an upper limit of flare energy averaged over the
CME of {$\lesssim$}{$4\times 10^{14}$} \ergg\@.  This quantity is
about an order of magnitude less than the kinetic energy per unit mass
derived from the LASCO CME catalog.

\subsection{\Yohkoh/SXT observations}

\Yohkoh/SXT \citep[][]{tsuneta:1991} performed pre-flare observations
including the precursor active region (AR 9046) from 12:00 UT until
13:54 UT and from 15:26 UT until 16:35 UT\@.  Post-flare
observations were taken from 19:26--19:46
UT\@.  All three data sets used the thin aluminum filter.  There were
no other bright X-ray sources in the SXT field of view during this
time.  
No observations were taken during the peak intensity phase of the
\GOES\ flare.  The observation at 16:50 UT shows that much of the soft
X-ray emission from the precursor active region was behind the limb.
The observations at 19:26 and 19:50 UT, however, show some new bright
X-ray emission in front of the limb.  Because of the gap in SXT
observations, we cannot rule out the possibility that the flare site
was behind the limb and thus partially occulted.  However, as shown
most recently by \citet{aarnio:2010}, it is possible to have powerful
and massive CMEs associated with relatively weak flares \citep[see
also][]{reeves:2005}.

\subsection {\SOHO/MDI observations}

\SOHO/MDI \citep{sch95} takes high spectral resolution images of the
\nickeli\ {$\lambda$}6768 absorption line to characterize velocity
oscillations and Zeeman splitting in the photosphere.  Full disk
magnetograms were taken every 96 minutes.  A subsection of the
magnetogram was taken to match the field of view and orientation of
the SXT observations.  We use level 1.8 line-of-sight magnetograms to
measure the magnetic field and then estimate the magnetic energy of
the active region prior to the event.  We have included in our
analysis magnetograms for 2000 June 25--28.  On 2000 June 28, the
flaring active region was on the limb.  Due to the lack of resolution
near the limb and in order to better characterize the magnetic energy
associated with the active region, we include in our analysis
magnetograms from 2000 June 25 when AR 9046 was on disk.


\section{FEATURE IDENTIFICATION AND CHARACTERISTICS}

\label{features}

In this section we identify six features observed by both UVCS and
LASCO with good density diagnostics.  The positions and UVCS
observation times of these features are shown in Table \ref{blobs},
along with the lines which were detected, undetected, or off of the
UVCS panels.  Additional properties of these features are shown in
Table \ref{oxygen}.

\begin{deluxetable*}{ccccccc}
\tabletypesize{\scriptsize}
\tablecaption{Features observed by {UVCS}\label{blobs}}
\tablewidth{0pt}
\tablehead{
\colhead{Blob} &
\colhead{Time} &
\colhead{P.A.} &
\colhead{$\rho/R_\odot$} &
\colhead{Observed lines} &
\colhead{Upper limits} &
\colhead{Off-panel}}
\startdata
\blobA & 19:12 & 276$^\circ$ & 2.45 & \ovi, \lya, \niii        & \cii, \ov]       & [\ov], \ciii, \lyb \\ 
\blobB & 19:14 & 287$^\circ$ & 2.35 & \ovi, \lya, \cii, \niiib & \ov],  \niiia & [\ov], \ciii, \lyb \\
\blobC & 19:16 & 283$^\circ$ & 2.37 & \ovi, \lya, \cii         & \ov],  \niii  & [\ov], \ciii, \lyb \\
\blobD & 19:21 & 284$^\circ$ & 2.36 & \ovi, \lya               &\ov],  \cii, \niii & [\ov], \ciii, \lyb\\
\blobE & 19:27 & 285$^\circ$ & 2.35 & \ovi, \ov], [\ov], \lya,   
\ciib &  \ciia, \niii & \ciii, \lyb          \\
\blobF & 20:11 & 294$^\circ$ & 2.32 & \ovi, \ov], [\ov], \ciii, \niiib  & \lya, \lyb, \cii, \niiia & ---
\enddata
\end{deluxetable*}

\begin{deluxetable*}{lccccccccc}
\tabletypesize{\scriptsize}
\tablecaption{Blobs Observed by UVCS\label{oxygen}}
\tablewidth{0pt}
\tablehead{
\colhead{} 
& \multicolumn{5}{c}{\ovi} 
& \multicolumn{2}{c}{\ov}
& \multicolumn{1}{c}{\lya}
& \multicolumn{1}{c}{LASCO}
\\
\colhead{Blob No.} &
\colhead{$I_{1032}/I_{1037}$} &
\colhead{$T_{\mathrm{max}}$} &
\colhead{$V_{LOS}$} &
\colhead{$V_{POS}$} &
\colhead{$\log n_e$} &
\colhead{$I_{1213}/I_{1218}$} &
\colhead{$\log n_e$} &
\colhead{$T_{\mathrm{max}}$} &
\colhead{$N_{e,\mathrm{max}}$} 
}
\startdata
\blobA & 1.72 & 30  &-759 & 1465\tablenotemark{c} &
7.0\tablenotemark{a} &\dots  &\dots  & 2.0 & 23 \\
\blobB & 2.21 &  24 &  -730 & 1656\tablenotemark{b} &
7.5\tablenotemark{a} &\dots  &\dots  & 2.5 & 19 \\
\blobC & 2.60 &  39  &-735 &  1654\tablenotemark{b}
&7.0\tablenotemark{a} & \dots & \dots & 2.5 & 22 \\
\blobD & 1.67 &  29  &-616 & 1531\tablenotemark{c} &
6.9\tablenotemark{a} & \dots &\dots  & 4.3 & 20 \\
\blobE & 3.73 &  8.7  & -470& 1748\tablenotemark{b} &
6.6\tablenotemark{a} & $0.16$--$0.5$ & $6.2$--$6.8$ & 0.9& 53 \\
\blobF & 2.14 &  20  & -331 & 250 &\dots  & $0.16$ & $6.8$\tablenotemark{a}
 &\dots  & 7.5 
\enddata
\tablecomments{Key quantities derived for each of the features
  presented in Section \ref{features}.  
  $I_{1032}/I_{1037}$ and $I_{1213}/I_{1218}$ are the line ratios used
  for the \ovi\ and \ov\ density diagnostics, respectively, described
  in Section \ref{diagnostics}.
  The line-of-sight and plane-of-sky velocities, \vlos\ and \vpos, are
  given in \kms\@.  The plane-of-sky velocities were derived using
  \ovi\ pumping information for blobs A--E and MK4 observations for
  blob F\@.
  The number densities derived from these diagnostics are in units of
  \cc\@.  
  $T_{\mathrm{max}}$ is the maximum temperature in MK for each species
  allowed from UVCS line widths.
  $N_{e,\mathrm{max}}$ is the maximum allowed column density from
  LASCO observations in units of $10^{16}$ cm$^{-2}$.
}
\tablenotetext{a}{Adopted density.}
\tablenotetext{b}{Indicates pumping of \ovia\ by {\lyb} at
  velocities near 1810 \kms.}
\tablenotetext{c}{Indicates pumping of \ovib\ by {\ovia} at
  velocities near 1650 \kms.}
\end{deluxetable*}

Blobs A and B correspond to the first clear detections of CME
prominence plasma at their respective positions along the UVCS slit.
There is strong \ovi\ and \lya\ emission, with probable detections of
\niii\@.  The large blueshifts of $759$ and $730$ \kms, respectively,
indicate that \lyb, \ciii, and [\ov] are off of the UVCS panel.  The
\ov] line is obscured by bright \lya, but upper limits of \ov] are
obtained.  The densities are found through radiative pumping of the
\ovi\ doublet to be {$\sim$}$1\times 10^7$ \cc\ for blob A and
{$\sim$}$3\times 10^7$ \cc\ for blob B\@.

Blobs C and D are located spatially between the locations of blobs A
and B along the UVCS slit but are observed by UVCS a few minutes
later.  The \lyb, [\ov], and \ciii\ lines are once again off-panel but
upper limits of \ov] are still possible despite bright \lya\@.  Again,
radiative pumping of \ovi\ is used to estimate number densities of
{$\sim$}$1\times 10^7$ \cc\ for blob C and {$\sim$}$8\times 10^6$ \cc\
for blob D\@.

Blob E is the first feature observed by UVCS with clear detections of
both [\ov] and \ov], allowing the density to be found using the
intensity ratio between these two lines.  The ratio of the \ovi\ lines
is $I_{1032}/I_{1038} = 3.73$, which is substantially different from
the collisional ratio of 2:1 and is indicative of significant
radiative pumping of the \ovi\ $\lambda 1032$ line.  Pumping of this
line by chromospheric \lyb\ indicates a total velocity of about 1810
\kms, which is large compared to $\vlos\approx -470$\ \kms\@.  The
density inferred from \ovi\ radiative pumping is $n_e\sim 4\times
10^6$ \cc.
This is within the range of densities derived from the \ov\ line ratio
of $1.5$--$6\times 10^6$ \cc\@.
This interpretation implies that $\vpos \gg
\vlos$, that this material is propagating close to the plane of sky,
and that this plasma was ejected perhaps $\sim$10--20 minutes after
the initial eruption.

A less likely possibility to explain the \ovi\ ratio observed in Blob
E is that the \ovi\ $\lambda 1032$ line was being radiatively pumped
by chromospheric \oi\ $\lambda\lambda 1027,1028$ emission at
velocities near 1100 and 1300 \kms.  According to the SUMER spectral
atlas of the solar disk presented by \citet{curdt:2001}, these two
lines are about an order of magnitude fainter than chromospheric
\lyb\@.  Consequently, this interpretation would imply that the number
density is an order of magnitude smaller than inferred using the
assumption of radiative pumping of \ovi\ $\lambda$1032 by \lyb\@.
Because the number density derived using the assumption of radiative
pumping by \lyb\ is consistent with the number density derived from
the \ov\ line ratio, we conclude that radiative pumping by \lyb\ is
much more likely.



Blob F is observed by UVCS much later in the event (20:11 UT).  We
identify Blob F as the rising filament structure observed earlier by
EIT and MK4\@.  This blob appears as a diagonal shear flow feature in
UVCS in \ov\ and \ovi\ emission.  There is \ciii\ emission with a
different morphology, suggesting that the cool gas is included at
least partially in a different component along the line of sight
\citep[see also][]{akmal:2001, lee:2009}.  \lya\ and \lyb\ emission
are both largely absent.  Consequently, both [\ov] and \ov] are easily
identified.  The \ov\ line ratio is used to derive a density
of $n_e \sim 6 \times 10^6$ \cc\@.

Column densities for each of these features are found using LASCO C2
observations rather than MK4 data because of the better signal to
noise near the UVCS slit position.  The mass per pixel for each C2
observation is found using the {\tt C2\_MASSIMG} function provided by
A.\ Vourlidas in SolarSoft IDL\@.  Because of the relatively low cadence of
C2 observations, the C2 position for each feature is estimated by
assuming that the plasma from each feature continues propagating on
the direct path between the flare site and the position on the UVCS
slit where it was observed.  The plane-of-sky velocity is assumed to
be constant and is found using blueshifts and \ovi\ pumping
information (for blobs A--E) or MK4 observations for (blob F).  {\tt
C2\_MASSIMG} assumes that the ejecta are propagating in the plane of
the sky so we use the ratio of \vpos\ and \vlos\ to correct for the
dilution factor and Thomson scattering angle.  We assume that the
column density in C2 observations goes down as $l^{-2}$, where $l$ is
the apparent distance from the flare site.  Because there will be some
departure from a constant velocity, we choose the largest column
density within an apparent radius of 0.1 to 0.3$R_\odot$, depending on
the estimated distance travelled between the UVCS and C2 observations.
The results are shown in Table \ref{oxygen} and will be used to
constrain the ionization fraction of \ovi\ in the following sections.

\section{METHOD}

\label{method}

To estimate plasma heating rates for the features described in Section
\ref{features}, we use a one-dimensional time-dependent ionization
code to track the ionization states of the plasma between the flare
site and when the features were observed by UVCS\@.  The numerical
method has been described in detail by \citet{akmal:2001} and
\citet{lee:2009}, and we discuss key features here.

Because we do not know the initial state of the plasma before the
eruption, we run a grid of models with different initial densities,
initial temperatures, heating rates, and heating parameterizations.
The initial densities are assumed to be in the range
$\log{\left(n/\mbox{cm}^{-3}\right)} \in [8.6,11.6]$.  The initial
temperatures are assumed to be in the range $\log
\left(T/\rm{K}\right) \in [4.6,6.6]$.  Consequently, the ejecta are
allowed to be cool prominence plasma at relatively high densities or
hot plasma from the ambient corona at lower densities.  The final
densities are presented in Table \ref{oxygen} and were found using one
or both of the diagnostics discussed in Section \ref{diagnostics}.
The determination of final density provides a significant constraint
on parameter space.  The velocity curve is scaled from the prominence
velocity curve for this event shown in Figure~2b of
\citet{maricic:2006}.  This velocity curve includes an acceleration
phase at low velocity.  The ejecta are assumed to be expanding
homologously.

We use four different heating parameterizations which we chose for
simplicity and consistency with previous work.  These simple
parameterizations allow us to explore parameter space and the spatial
dependence of heating.
The first parameterization is the wave heating model for the fast
solar wind presented by \citet{allen:1998}, denoted by
$Q_{\mathrm{AHH}}\propto \exp\left(-d/0.7R_\odot\right)$, where $d$ is
height above the limb.
This is physically motivated by wave heating models of the solar
corona.  While there may be departures from an exponential if, for
example, the Alfv\'en speed changes with altitude, this form is a
reasonable approximation that allows for a gradual decrease of heating
with height that does not depend on density or lead to
excessive heating far from the flare site.
The second parameterization is heating proportional to number density,
$Q \propto n$, which also gives strong heating at low heights with a
gradual decrease at higher altitudes.
The third parameterization is heating proportional to the number
density squared, $Q\propto n^2$, which concentrates heating at low
heights where the density is large and has the same density dependence
as radiative cooling.  No physical mechanism is assumed for $Q\propto
n$ and $Q\propto n^2$.  A compilation of heating models for coronal
loops is available in Table 5 of \citet{mandrini:2000} with some
showing heating proportional to a power of density, but it is not
clear how applicable these models are to CMEs\@.
The fourth parameterization is heating proportional to the time
derivative of the sum of the kinetic and gravitational energies,
$Q_{\mathrm{KR}}\propto \dif\!\left(U_\mathrm{KE} +
U_\mathrm{G}\right)/\dif t$.  This expression arises from Eq.\ 71 of
KR when we assume for simplicity that a constant fraction of the
released magnetic energy goes into heating the plasma within each
model (e.g., $\dif Q = -h\,\dif U_\mathrm{m}$ where $h$ is the
fraction of lost magnetic energy appearing as heat which we assume to
be a model parameter that is constant within each run and
$U_\mathrm{m}$ is the magnetic energy).\footnote
{
  As pointed out by the referee of this paper, Eq.\ 72a of KR with
  $\beta_G\cong 0.22$ implies that Eq.\ 72b should be $0.14 \leq h
  \leq 0.78$.  This lower limit of 14\% of the magnetic energy 
  available for heating is in contrast to the lower limit of 58\%
  reported by KR.  We were able to reproduce Eq.\ 72a of KR using the
  expressions $\dif U_{KE}/\dif U_m = - 2(1-s)$ and $\beta_G = - \dif
  U_G/\dif U_\mathrm{m}$.  The latter expression suggests that
  $\beta_G$ and therefore $h$ will in general be functions of time
  except when $\dif U_\mathrm{m}/\dif t \propto -\dif U_G/\dif t$ (or, 
  if we assume a point mass expression for gravitational energy,  
  $\dif U_\mathrm{m}/\dif t \propto -V/R^2$).
}

For each model the ionization states are evolved using the relation
\begin{eqnarray}
  \frac{\dif n_z}{\dif t} &=& \nonumber
  n_e n_{z-1} q_i(Z,z-1,T) \\ & &
  - n_e n_z \left[ q_i(Z,z,T) + \alpha_r(Z,z,T) \right] \nonumber \\ & &
  + n_e n_{z+1} \alpha_r(Z,z+1,T),
\end{eqnarray}
where $q_i(Z,z,T)$ and $\alpha_r(Z,z,T)$ are the ionization and
recombination rate coefficients for an ion $z$ of element $Z$ at a
temperature $T$.  Initially, each model is in ionization equilibrium
for the assumed starting temperature with coronal abundances.  The
particle distribution functions are assumed to be Maxwellian.  The
elements considered in this analysis are H, He, C, N, O, Ne, Mg, Si,
S, Ar, Ca, and Fe.  Cooling by radiative losses and adiabatic
expansion are included in the analysis.  A temperature floor is
maintained at 5000 K, and model runs are rejected when the temperature
exceeds $10^7$~K\@.  

Once the grid of models is completed for each feature, the predicted
line intensities are compared to UVCS observations to determine which
sets of parameters are acceptable.  The emission is assumed to come
from the features bright in \ov\ and \ovi, while all other lines are
used as upper limits since they might be emitted from a different
component of the plasma along the line of sight.  The UVCS
observations alone constrain the line ratios, but the LASCO column
densities constrain the ionization fraction of \ovi, thus allowing us
to place limits on the absolute line strengths.  Line widths for \ovi\
and \lya\ provide (usually uninteresting) upper limits on the final
temperature.  The results of this analysis are discussed in the
following section.

\section{CONSTRAINTS ON PLASMA HEATING DERIVED FROM UVCS
  OBSERVATIONS} 

\label{results}

Constraints on plasma heating for the features described in Section
\ref{features} derived using the techniques described in Section
\ref{method} are presented in Table \ref{results_table}.  This table
shows the components of the energy budget in terms of energy per unit
mass, including the kinetic energy, gravitational potential energy,
cumulative heating energy, and thermal energy of the plasma when
observed by UVCS\@.

\begin{deluxetable*}{lcccccccccc}
\tabletypesize{\scriptsize} 
\tablecaption{Energy Budgets for Multiple Heating Models in Units of
  $10^{14}$ \ergg\label{results_table}} 
\tablewidth{0pt} \tablehead{ 
\colhead{} & \colhead{} & \colhead{} & \multicolumn{2}{c}{$Q \propto
Q_{\mathrm{AHH}}$} & \multicolumn{2}{c}{$Q \propto n$} &
\multicolumn{2}{c}{$Q \propto n^2$} & \multicolumn{2}{c}{$Q \propto
Q_{\mathrm{KR}}$} \\ \colhead{Blob No.} & \colhead{K.E.} &
\colhead{G.E.} & \colhead{H.E.} & \colhead{T.E.} & \colhead{H.E.} &
\colhead{T.E.} & \colhead{H.E.} & \colhead{T.E.} & \colhead{H.E.} &
\colhead{T.E.}
}
\startdata
\blobA & 136 ($>$29) & 7.4--7.8 & 5.5--34.5  & 0.62--4.1 & 7.2--45.6  & 0.31--2.0 & 22.3--42.0 & 0.03--0.05 & 7.4--127   & 0.10--1.6 \\
\blobB & 164 ($>$27) & 7.9--8.1 & 0.26--36.6  & 0.03--4.8 & 1.4--85.8   & 0.07--4.25 & 18.4--117   & 0.03--0.21& 6.6--379    & 0.1--4.6 \\
\blobC & 164 ($>$27) & 7.7--8.1 & 0.15--35.5   & 0.02--4.6 & 0.55--87& 0.03--4.3 & 11.9--112    & 0.02--0.2    & 1.3--392   & 0.02--4.7 \\
\blobD & 136 ($>$19) & 7.9--8.1 & 0.17--60.5 & 0.02--7.5 & 0.41--163 & 0.02--7.4 & 13.3--112 & 0.02--0.15& 1.4--422   & 0.02--5.6 \\
\blobE & 164 ($>$11) & 8.2--8.3 & 1.6--12.6 & 0.2--1.6 & 3.3--13.1 & 0.16--0.64 & 17.4--109.4 & 0.03--0.2 & 5.9--29.8 & 0.09--0.44 \\
\blobF & 8.6  ($>$5.5) & 5.5--8.2 & 6.5--8.2 & 0.67--0.95 & 16.9 & 0.9 & --- & --- & 56.6 & 0.41 
\enddata
\tablecomments{
  Components of the energy budget for the features observed by UVCS
  during this event, including the kinetic energy (K.E.),
  gravitational potential energy (G.E.), the cumulative heating energy
  (H.E.), and the thermal energy, shown
  for each of the heating parameterizations described in Section \ref{method},
  assuming 10\% helium.
  The kinetic energy is estimated using the velocities of \ovi\
  pumping for blobs A--E and MK4 observations for blob F, with the
  lower limits on the kinetic energies for all features found using
  \vlos\@.
  The gravitational potential energy is given by $GM_{\odot}/R$
  where $R$ ranges from the apparent radius in the plane of the sky
  and the deprojected radius. 
}
\end{deluxetable*}

\begin{figure}
  \begin{center}
  \includegraphics[scale=1.0]{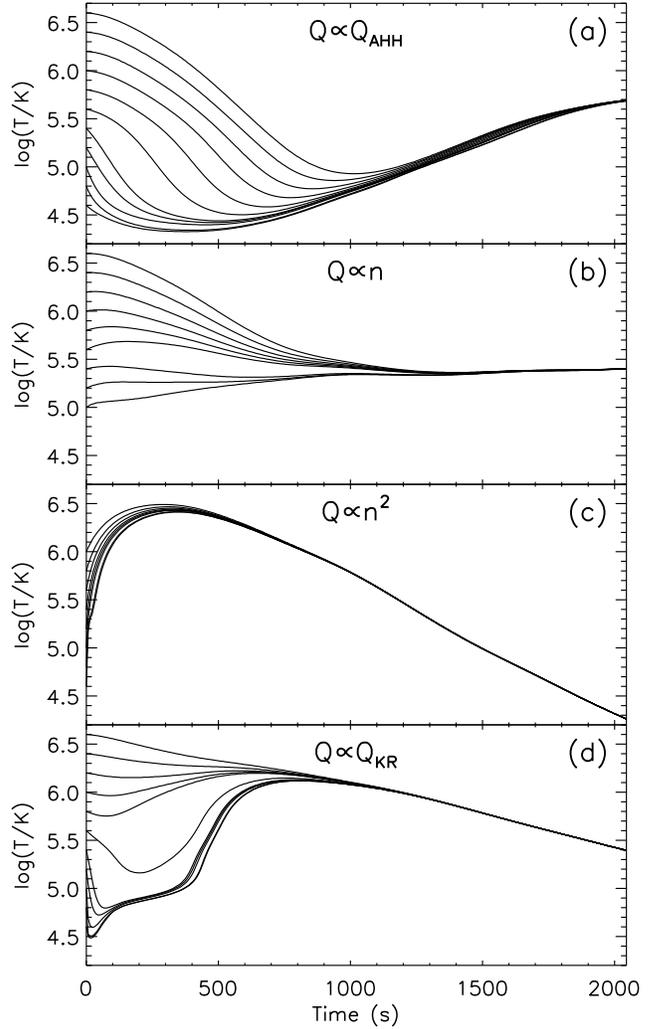}
  \end{center}
  \caption{Characteristic temperature histories for each of the
    different heating mechanisms shown using selected models for blob
    \blobA\@.  \label{blobAfig}}
\end{figure}

The different heating parameterizations typically each have a
characteristic temperature history pattern, as shown in Figure
\ref{blobAfig}.  For example, the allowed wave heating models
($Q_{\mathrm{AHH}}$) typically show the temperature dropping to
$\lesssim${$10^5$} K before gradually heating up to a few times $10^5$
K at the time when observed by UVCS (Figure \ref{blobAfig}a).
Heating proportional to number density ($Q\propto n$) generally leads
to a steady temperature after the ejecta leave the vicinity of the
flare site (Figure \ref{blobAfig}b).  Occasionally, the temperature
does approach the floor value for this parameterization during the
middle of a run also.  When $Q\propto n^2$, the ejecta tend to be
heated to up to a few times $10^6$~K before gradually adiabatically
cooling to temperatures below {$10^5$~K} (Figure \ref{blobAfig}c).
The heating parameterization by KR does not greatly constrain the
temperature history at early times but eventually tends to result in a
gradually decreasing temperature at later times where heating does not
provide enough energy to completely counter adiabatic cooling and
radiative losses (Figure \ref{blobAfig}d).

There are interesting upper limits on cumulative plasma heating for
all features, and interesting lower limits on heating for blobs A and
F (and to a lesser extent, blob E).  For blobs A and E, the cumulative
plasma heating is constrained to be less than the inferred kinetic
energy of each of the features.  For blobs B--D, the cumulative plasma
heating is constrained to be less than {$\sim$}$2$--$3$ the inferred
kinetic energy.  For blob F, the slowest feature, the plasma heating
is constrained to be comparable to or greater than the inferred
kinetic energy.

The results for blob E, the fast feature observed slightly later
during the event, indicate that significant plasma heating comparable
to the kinetic energy could only have occured early in the event when
the ejected plasma was not far from the flare site.  Heating
proportional to the square of the density does allow heating up to
$109\times 10^{14}$ \ergg, but for this mechanism the heating is
strongly concentrated at low heights where the density is high.  The
other parameterizations which allow for more gradual heating do not
allow the cumulative heating to be greater than $\sim${$0.2$} of the
kinetic energy for this feature.

\begin{figure}
  \begin{center}
  \includegraphics[scale=1.0]{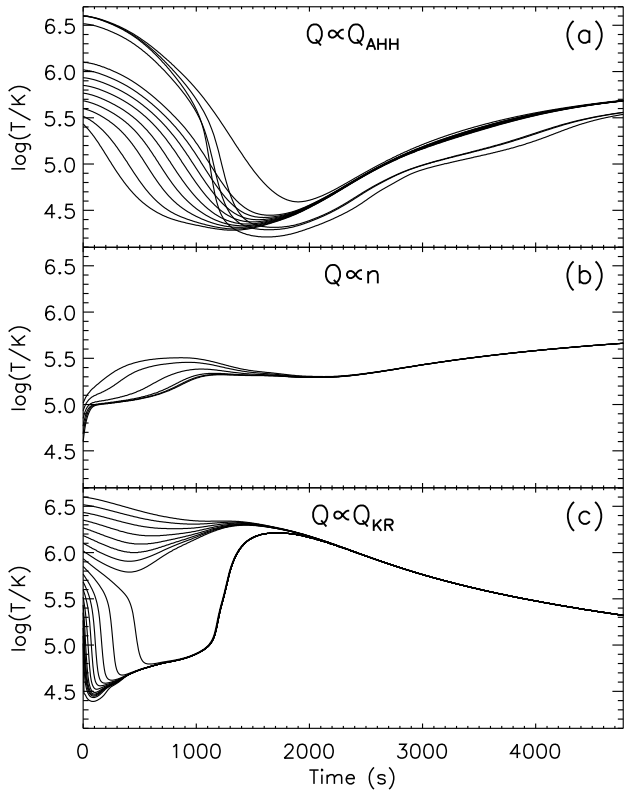}
  \end{center}
  \caption{Temperature histories for the allowed models for blob F\@.
  \label{blobFfig}}
\end{figure}

Blob F, the slow feature observed late in the event, has the best
constraints because it is the only feature for which the observed
\ciii\ line was completely on the UVCS panel.  
For $Q\propto Q_{\mathrm{AHH}}$, the temperature drops below {$5\times
10^4$} K for all of the allowed runs (Figure \ref{blobFfig}a).  This
behavior is possible but unlikely, and was ruled out by
\citet{landi:2010} for a separate event.
For $Q\propto n$, the allowed runs show steady or slowly increasing
temperature histories at a few times $10^5$ K (Figure
\ref{blobFfig}b).  The initial densities are $4\times 10^8$ \cc, which
is a coronal rather than prominence density at the low end of our
assumed density distribution.  However, the initial temperature is
constrained to be {$<$}$10^5$ K for this parameterization, which would
be unusual for plasma at coronal densities.
The allowed models with $Q\propto Q_{\mathrm{KR}}$ show a gradually
decreasing temperature history with initial densities of
$10^{10}$~\cc, a broad range of initial temperatures, and a final
temperature of $2\times 10^5$ K (Figure \ref{blobFfig}c).

\section{ACTIVE REGION MAGNETIC ENERGY}

\label{magneticenergy}

We estimate the total magnetic energy of the pre- and post-CME active
region (AR 9046) using \SOHO/MDI observations.
We use an empirical relationship between
the total unsigned magnetic flux and the magnetic energy,
\begin{equation}
  E_{\mathrm{mag}} \simeq 2\times 10^{32} 
  \left(
    \frac{\Phi_{\mathrm{tot}}}{10^{22}\mbox{~Mx}}
  \right)^{1.35}\mbox{~ergs},
  \label{mag_energy_mdi}
\end{equation}
presented by \citet{fis98}.  The coefficient for Equation
\ref{mag_energy_mdi} is estimated using Figure 8b of
\citet{fis98}. This method assumes that the magnetic field is
potential and the results should thus be considered approximate lower
limits or order of magnitude calculations.  It should be mentioned
that the magnetic free energy is not accounted for but is probably of
the same order of magnitude as the magnetic field energy found by
assuming a potential field.



Using the data set of magnetograms for 2000 June 25 and Equation
\ref{mag_energy_mdi} we obtain an average magnetic energy of $8.6
\times 10^{31}$ ergs.  Averaged over the day on 2000 June 28, the
magnetic energy is found to be $1.8 \times 10^{31}$ ergs but is very
uncertain because of the active region's proximity to the limb.  The
magnetic energy estimates immediately before and after the event do
not significantly vary; this is expected because magnetic free energy
is used to power the CME, and the CME itself should not significantly
modify the line-tied magnetic field near the photosphere.  Using the
LASCO CME Catalog's (uncertain) representative mass with the 2000 June
25 magnetic energy estimate, this corresponds to {$\sim$}{$110\times
10^{14}$} \ergg\ and is comparable to the representative kinetic
energy of the event.





\section{CANDIDATE HEATING MECHANISMS}

\label{mechanisms}

The most probable source of plasma heating during CMEs is the magnetic
field.  However, the mechanism which dissipates magnetic energy into
thermal energy is not understood.  In this section, we discuss several
candidate mechanisms which could be responsible for heating the
ejected plasma.  While quantitative model predictions for these
mechanisms are generally not yet available, we can provide
observational constraints on some of these proposed mechanisms for
this event.

\subsection{Heating by the CME current sheet}\label{cmecs}

Flux rope models of CMEs predict the formation of an elongated current
sheet in the wake of the rising plasmoid
\citep[e.g.,][]{kopp:pneuman:1976, linforbes:2000}.  Features
identified as current sheets have been observed during a number of
CMEs \citep[e.g.,][]{ciaravella:2002, ciaravella:2008, schettino:2010,
savage:2010, patsourakos:2011, reeves:2011}.
These current sheets are expected to be unstable to the formation of
plasmoids which may facilitate fast reconnection
\citep[][]{loureiro:2007, samtaney:2009, bhattacharjee:2009,
huang:2010:plasmoid, shepherd:2010, ni:2010, uzdensky:2010,
barta:2008, barta:2011a}.
These current sheets have the potential to increase the thermal energy
content of CMEs via antisunward-directed exhaust.  Recent analytical
\citep{seaton:2008, murphy:asym} and numerical \citep{reeves:2010,
murphy:retreat} studies predict that most of the mass, momentum, and
energy flux from the CME current sheet will be directed upward towards
the rising plasmoid.  \citet{lin:2004} argue that current sheet
exhaust piles up in and around the rising flux rope so that the final
parameters for CME evolution are set at heights of several solar
radii.

Outflow from the CME current sheet has the potential to account for
two key observed features of CMEs: that CMEs continue to be heated far
from the flare site, and that the masses of CMEs tend to increase with
time at distances of up to {$\sim$}$3$--$10R_{\odot}$ from the Sun
\citep{bemporad:2007, vourlidas:2010}.  However, current observations
do not strongly constrain the mass or energy budgets of CME current
sheets.  In particular, it is not known whether CME current sheets
contribute substantially to the mass and energy budgets of the CME as
a whole as some simulations and theories suggest
\citep[e.g.,][]{lin:2004, reeves:2010}.

The predictions of the CME current sheet heating mechanism depend
strongly on the importance of mixing and transport.  In the model by
\citet{lin:2004}, transport of reconnection outflow around the flux
rope is assumed to fill each newly reconnected flux surface quickly
without mixing plasma from different flux surfaces.  However, the
reconnection outflow jets could penetrate deep into the flux rope and
lead to significant turbulence and mixing (J.\ Karpen, private
communication, 2010).  Mixing acts to spread the effects of heating
over a larger volume.  If mixing effects are not important, then the
\citet{lin:2004} model predicts reconnection heated plasma surrounding
a core of cool material.  \citet{shiota:2005} present simulations
which suggest that slow shocks can permeate the flux rope, thus
further contributing to heating.  Further quantitative predictions of
this heating mechanism can be made using a combination of numerical
simulations and analytic theory.

Cool cores are commonly observed \citep[e.g.,][]{akmal:2001},
suggesting that the cool plasma may be magnetically isolated from
hotter nearby plasma.  Some events do show significant heating in the
core of the CME \cite[e.g.,][]{bemporad:2007}.  \Insitu\ measurements
of low ionization state plasma in ICMEs \citep[e.g.,][]{lepri:2010}
also provide constraints on mixing below heights where ionization
freezes in.  However, observations of the 2000 June 28 CME do not
provide significant constraints on the efficacy of heating by the CME
current sheet.

Extending the analysis performed in this paper to a large number of
events will provide some information on the spatial dependence of CME
heating.  However, none the heating parameterizations described in
Section \ref{method} directly correspond to heating of ejecta by the
CME current sheet.  Numerical simulations such as those by
\citet{reeves:2010} can be used to investigate the efficacy of this
mechanism, especially when used in conjunction with a time-dependent
ionization code to facilitate a comparison to observations.

\subsection{The kink instability}\label{kinking}

The kink instability is driven by current parallel to the magnetic
field and causes long flux ropes to develop a characteristic helical
shape.  Kinking is frequently observed in prominences and during solar
eruptions \citep[e.g.,][]{rust:2005}.  Line-tying has a stabilizing
effect on the kink instability \citep{huang:2006,
huang:linetying:2010}, but this mode may be destabilized by flux rope
curvature.  This instability has the potential to heat CME plasma by
injecting turbulence through large scale motions.  The turbulence then
dissipates as the CME propagates away from the Sun.  If this mechanism
is most important, then it is likely that heating would occur over a
turbulent dissipation time scale and would be greater in the vicinity
of the flux rope.

LASCO and MK4 observations show that the flux rope becomes twisted.
This behavior is consistent with the kink instability, but could be
due to different effects.  However, the transverse motions associated
with this kink instability are much slower than the outward
propagation of the flux rope.  Thus the kink instability, by itself,
is not likely to release enough energy through mass motions to explain
total heating comparable to or greater than the total kinetic energy
of a blob, as observed for several features during this and other
events.
However, additional magnetic energy could be released through
reconnection events driven by the kinking behavior.  This process is
analogous to the kink-tearing behavior during sawteeth in tokamaks and
other laboratory plasma confinement devices.

\subsection{Small-scale magnetic reconnection}\label{smallscale}

The candidate heating mechanisms described in Sections \ref{cmecs} and
\ref{kinking} are intrinsically linked to large-scale CME dynamics.
An alternative to these models is heating through small-scale,
three-dimensional reconnection events which might or might not be
driven by global dynamics.  For example, flux rope expansion models
such as those by KR and \citet{wang:2009} predict that a significant
fraction of the magnetic energy is converted into kinetic and thermal
energy.  The mechanisms by which this process occurs are not specified
for these models but are very likely to be some form of magnetic
dissipation (likely through some combination of reconnection and
turbulence).  
\citet{owens:2009} presents a model which shows how internal
reconnection can occur as a natural result of flux rope expansion 
\citep[see also][]{xu:2011}.
Small-scale reconnection is probably needed for CMEs to relax from a
complex structure to the Lundquist configurations often observed in
interplantary magnetic clouds \citep[e.g.,][]{lynch:2004}.
Heating by small-scale reconnection is analogous to the nanoflare
model of coronal heating.  
One possible manifestation of this candidate heating mechanism is the
tearing mode \citep{fkr}.  The tearing mode occurs frequently in
magnetically confined laboratory plasmas in a variety of
configurations and can moreover be driven by the kink instability.  As
in the case of the kink instability, line-tying where the flux rope is
attached to the photosphere provides a stabilizing effect and can
change the eigenmode structure and how this mode scales with
resistivity \citep{huang:2009, huang:linetying:2010}.  Tearing
behavior can give rise to turbulence which can behave as an effective
hyperresistivity \citep[e.g.,][]{vanballegooijen:2008, strauss:1988}.

Because these processes occur on small-scales and magnetic fields
during CMEs are very difficult to diagnose, there are few
observational constraints for this candidate mechanism.  During the
2000 June 28 CME, the \ovi\ widths in blob F are $\approx${$100$}
\kms\@.  This corresponds to an upper limit on the turbulent energy
density of {$\sim$}$0.5\times 10^{14}$ \ergg, although there is some
contribution to the line width from thermal broadening and perhaps
shear flow.  This upper limit indicates that turbulence must be
continually ejected into the system and dissipated on time scales much
shorter than the CME propagation/expansion time scale.  Observational
signatures of this mechanism include Alfv\'enic outflows and heating
concentrated near the regions of maximum shear in the flux rope.

\subsection{Wave heating}

One of the leading mechanisms for heating of the ambient corona and
active regions is the damping of MHD waves driven by photospheric
motions.  The exponential heating parameterization used in this
analysis was derived in the context of wave heating.  We find that
heating rates of $\gtrsim${$100$} and $\gtrsim${$63$} times the
coronal hole heating rate of \citet{allen:1998} are required for blobs
A and F, respectively, to explain the observed emission with this
heating mechanism.  This is consistent with the model behavior
observed by \citet{landi:2010}, who showed that heating rates
$\gtrsim${$1500$} times the \citet{allen:1998} heating rate were
required with this model during the 2008 April 9 CME\@.  For blob F,
each of the allowed wave heating models shows a drop in temperature
{$\lesssim$}$5\times 10^4$ K, which is possible but unlikely.  These
results provide further evidence that dissipation of MHD waves driven
by photospheric motions is not able to explain the heating of CME
plasma.

Alternatively, recent laboratory experiments of the eruption of arched
magnetic flux ropes show that intense fast magnetosonic waves
resulting from the eruption are capable of tranferring energy to and
heating the plasma in and around the erupting flux rope
\citep{tripathi:2010}.  This mechanism might occur in CMEs, possibly
in conjunction with a large scale instability of the flux rope.  The
observations of the 2000 June 28 CME do not provide meaningful
constraints for this mechanism, except for the upper limit on
turbulent energy density at UVCS heights discussed in Section
\ref{smallscale}.

\subsection{Thermal conduction}

Thermal conduction along magnetic field lines is quick and therefore a
potential contributor to the heating of CME plasma.  The efficacy of
this mechanism is limited, however, because of the short CME
propagation time scales and long length scales.  \citet{landi:2010}
examine thermal conduction in the 2008 April 9 CME and find that
heating due to conduction is not sufficient to explain the observed
heating rates.

As discussed in Section \ref{goesobservations}, the 2000 June 28 CME
is associated with a relatively weak C class flare.  Assuming that the
CME is heated approximately uniformly and that the flare was not
strongly occulted by the solar limb, the lower limits on plasma
heating for blobs A and F suggest that essentially all of the flare
energy available must go into CME heating for this mechanism to be
important.  Therefore, it is unlikely that thermal conduction between
the flare site and the ejecta is responsible for the inferred plasma
heating.

\subsection{Energetic particles}

Some contribution to CME heating may be due to energetic particles
that were accelerated during the impulsive phase of the event.  A
clear understanding of this mechanism requires a detailed description
of the deposition of energy by energetic particles into the bulk
plasma \citep[see, for example,][]{allred:2005}.  The efficacy of this
mechanism may be limited because the particle acceleration phase of
flares is short compared to CME propagation time scales, and we know
that heating continues at large distances from the flare site for many
events.

If the flare associated with the 2000 June 28 CME was not behind the
limb, then it would be reasonable to infer that relatively few
energetic particles were accelerated during the impulsive phase of
this event.  Additionally, blob F was observed {$\sim$}$80$ minutes
after flare onset which is probably too late for energetic particles
accelerated from the flare to do much heating.  Thus we conclude that
energetic particles from the flare site are not likely to be
responsible for substantially heating the ejecta.

One complicating factor of energetic particles for this form of
analysis is that non-Maxwellian particle distributions can increase
the ionization rates substantially in addition to heating the thermal
component of the plasma.  It would be interesting in future work to
see how a non-Maxwellian tail in the particle distribution function
could affect the inferred heating rates.

\subsection{Heating by counteracting flows}

\citet{filippov:2002} describe a heating mechanism for erupting
prominences where upward concave flux rope segments yield shocks from
colliding flows accelerated by gravity.  Because of the dependence on
gravity, this mechanism is limited by the amount of gravitational
energy available unless the plasma drops and rises multiple times.
Between $1.1$ and $3.0R_\odot$, there is a difference in gravitational
potential energy of $11\times 10^{14}$ \ergg.  This is much less than
the kinetic energies of most of the features but greater than the
lower limits on plasma heating for all of the features.  For blob F,
this is true only for the wave heating mechanism.  However, the
efficiency of this process in terms of available gravitational energy
is probably less than unity since the distance plasma falls is likely
to be less than the total distance available, and only a fraction of
the plasma in a flux rope is likely to fall.  Thus, unless the ejected
plasma rises and falls multiple times, we conclude that this mechanism
is unlikely to account for sufficient heating during this event.

\subsection{Ohmic heating from net current in the flux rope}

The model by \citet{chen:1996} suggests that the injection of magnetic
flux into a flux rope is a trigger for CMEs and that there is a net
current through the flux rope.  Assuming a characteristic current of
{$\sim$}$10^{11}$~A \citep[e.g.,][]{chen:1996}, a flux rope minor
radius of {$\sim$}$0.25R_\odot$, a propagation time of
{$\sim$}$10^3$~s, and an electrical diffusivity of $\eta\sim 10^4$
cm$^2$~s$^{-1}$, we derive an expected heating rate of {$\sim$}$10^5$
\ergg\@.  It is not surprising that this estimate is nine orders of
magnitude below the lower bounds on blobs A and F, and thus cannot
explain the inferred heating.

The above estimate makes two implicit assumptions.  First, the current
density is assumed to be roughly uniform across the flux rope cross
section.  Alternatively, the current density could be very strongly
concentrated, which would indicate small-scale reconnection phenomena
(see Section \ref{smallscale}).  Second, we assume classical Spitzer
resistivity.  An anomalous resistivity due to wave-particle
interactions may be present, but would need to be many orders of
magnitude larger than the Spitzer resistivity
\citep[e.g.,][]{rakowski:2011}.  Anomalous resistivity generally
requires large currents or sharp gradients over short length scales
(i.e., the ion inertial length or ion sound gyroradius).  Again,
strong current filamentation would be necessary and this would
indicate small-scale reconnection phenomena as a more relevant model.
Moreover, it is not clear whether the processes which lead to
anomalous resistivity would produce volumetric heating best described
as being proportional to the square of the the current density.  
For these reasons we conclude that resistive heating from net current
in the flux rope is not a viable means of adequately heating CME
plasma.

\section{CONCLUSIONS}\label{conclusions}

In this paper we use a one-dimensional time-dependent ionization code
to investigate the energy budget of a CME observed by \SOHO/UVCS\@.
By running a grid of models with different initial densities, initial
temperatures, and heating parameterizations, we are able to constrain
the total amount of heat deposited into the ejected plasma to counter
radiative losses and adiabatic cooling.  

We perform this analysis for six features observed by UVCS and
LASCO\@.  Number densities of the plasma observed by UVCS are found
either through radiative pumping of the \ovi\ $\lambda\lambda
1032,1038$ doublet \citep[e.g.,][]{raymond:2004,noci:1987} or by the
classical density sensitive \ov\ $\lambda\lambda 1213,1218$ doublet
\citep[e.g.,][]{akmal:2001}.  Both of these diagnostics are available
for one feature (blob E) and the derived number densities are
consistent to within the expected systematic errors.  Total velocity
information is found using radiative pumping of the \ovi\ doublet and
through white light observations.

For two of the features (blobs A and E), the cumulative plasma heating
is constrained to be less than or comparable to the inferred kinetic
energy of the feature when observed by UVCS\@.  For three of the
features (blobs B--D), the cumulative heating is constrained to be
less than {$\sim$}$2$--$3$ times the inferred kinetic energy.  For the
slow feature observed late in the event (blob F), the plasma heating
is constrained to be comparable to or greater than the kinetic energy.
Lower limits from two of the features (blobs A and F) yield cumulative
heating energies of {$\gtrsim$}$5\times 10^{14}$ \ergg\@.

Next we discuss and consider constraints on a variety of heating
mechanisms.  Upflow from the current sheet that forms in the wake
behind the rising flux rope could contribute substantially to both the
mass and energy budgets of CMEs, but the energetics of CME current
sheets are not well constrained for this or other events.  The kink
instability is able to drive turbulence by twisting of the flux rope,
but the observed motions do not contain enough energy to heat the
plasma for this event.  Secondary reconnection or tearing behavior
(perhaps driven by the kink instability) can drive turbulence which
dissipates and heats the plasma.  Wave heating can occur either
through photospheric motions \citep[cf.][]{landi:2010} or by waves
generated by the eruption of the flux rope itself
\citep[e.g.,][]{tripathi:2010}.  As also shown by \citet{landi:2010},
wave heating by photospheric motions is unlikely to be important for
CMEs since the required heating rates are orders of magnitude larger
than those inferred from the observed nonthermal mass motions.
Thermal conduction can bring in thermal energy from the flare site or
perhaps from the ambient corona.  Energetic particles could deposit
energy into the ejecta, but also increase ionization rates which would
complicate this analysis.  Because this event was associated with a
weak (C class) flare, we consider heating by thermal conduction or
energetic particles unlikely to be important; however, the flare might
have been partially occulted by the solar limb.  \citet{filippov:2002}
suggest that heating could be due to colliding flows which were
accelerated by gravity, but we conclude this is unlikely for this
event since lower limits on heating for several features are
comparable or greater than the gravitational energy available
\citep[see also][]{landi:2010}.


We estimate the magnetic energy of the precursor active region using
\SOHO/MDI observations and an empirical relationship by \citet{fis98}.
This estimate is uncertain because the active region is near the limb,
but a few days before the event the magnetic energy is estimated to be
$\sim${$8.6\times 10^{31}$} ergs.  This is the same order of magnitude
as the representative kinetic energy from the LASCO CME catalog.  This
estimate assumes a potential magnetic field and thus is probably an
underestimate.


%

There are likely to be selection effects associated with the sample of
events studied with the technique used in this paper
\citep{akmal:2001, ciaravella:2001, lee:2009, landi:2010}.  The UVCS
density diagnostics most often used are spectral lines of \ov\ and
\ovi\ which are most prevalent in plasmas at temperatures of order
{$\sim$}$10^5$~K and most useful at number densities between $10^6$
and $10^7$ \cc\@.  Additional diagnostics available for CME cores in
the lower corona include density sensitive line ratios of \oiv,
\mgvii, and \feviii\ and are accessible with instruments such as the
EUV Imaging Spectrometer \citep[EIS;][]{culhane:2007} on \Hinode.
Thus with these analyses we miss plasma at cooler or hotter
temperatures.  For this event, we do not see morphological features in
MK4 or LASCO observations which do not have UVCS counterparts, but
there may be additional components along the line of sight.

Thus far this form of analysis has been used to constrain CME heating
rates one event at a time.  However, the model assumptions made by
this and previous works differ slightly, thus complicating attempts to
make a systematic or statistical analysis of the problem.  In future
work, we will perform a standardized time-dependent ionization
analysis for features in a large number of events observed by
\SOHO/UVCS\@.  This will allow us to make direct comparisons of plasma
heating rates during different events and provide tighter constraints
on several mechanisms for CME heating.  High cadence observations by
the Atmospheric Imaging Assembly (AIA) on the \emph{Solar Dynamics
Observatory} (\emph{SDO}) are also well-suited for this analysis when
appropriate density diagnostics are available.  However, more detailed
model predictions will be needed before most of the candidate
mechanisms could be definitively ruled out.

In particular there are several open questions pertaining to the
energetics of CMEs.  These include: (1) What physical mechanisms are
most responsible for heating CME plasma?  (2) Are CME current sheets
energetically important to CMEs as a whole?  (3) How uniform is
heating within a CME? (4) How does CME heating depend on global CME
properties such as speed, mass, and magnetic field strength and
configuration? (5) What are the roles of energetic particles?  (6) How
do the kinetic energy and cumulative heating energy compare to a CME's
magnetic energy?  (7) How do CME and ICME flux ropes ``relax''
\citep[cf.][]{taylor:1986}? (8) Is magnetic helicity conserved during
the evolution of these systems?  We will address these questions in
future work.

\acknowledgments

The authors acknowledge useful discussions with 
J.-Y.\ Lee, 
M.~P.\ Miralles,
K.~K.\ Reeves,
S.\ Cranmer,
J.\ Lin,
A.\ Vourlidas,
A.~A.\ van Ballegooijen, 
H.\ Johnson,
E.~G.\ Zweibel,
T.~G.\ Forbes, 
J.\ Chen,
and
E.\ Robbrecht.  
We thank an anonymous referee for particularly insightful comments
which helped to improve this paper.
This research is supported by NASA grant NNX09AB17G to the Smithsonian
Astrophysical Observatory.  
\SOHO\ is a project of international cooperation between ESA and NASA.
MLSO is a facility of the National Center for Atmospheric Research
operated by the High Altitude Observatory.
\Yohkoh\ is a project of the Institute of Space and Astronautical
Science of Japan.
The LASCO CME catalog is generated and maintained at the CDAW Data
Center by NASA and The Catholic University of America in cooperation
with the Naval Research Laboratory.
This work has benefited from the use of NASA's Astrophysics Data
System.

{\it Facilities:} \facility{\SOHO\ (UVCS, EIT, LASCO, MDI)},
\facility{MLSO (MK4)}, \facility{\Yohkoh\ (SXT)}, \facility{\GOES}




\end{document}